\documentclass{article}
\usepackage{graphicx}

\begin{document}

\begin{titlepage}

\Large
\centerline{\bf CKM matrix: the `over-consistent' picture of the unitarity triangle~\footnote{Contribution to Beauty 2000, September 13-18 2000, Kibbutz Maagan, Israel. To appear in the Proceedings (Nucl. Instr. Meth. A).}}
\normalsize

\vskip 3.0 cm
\centerline {Pietro Faccioli}
\centerline {\it Dipartimento di Fisica dell'Universit\`a di Bologna and INFN Sezione di Bologna}
\centerline {\it E-mail: pietro.faccioli@libero.it, faccioli@bo.infn.it}
\vskip 4.0cm

\centerline {\bf Abstract}
\vskip 1.0cm

In presenting an up-to-date account of the experimental knowledge of the CKM matrix, special emphasis is placed on the exceptional degree of consistency shown by the current Standard Model determination of the unitarity triangle; some implications in the question of how the theoretical nature of the dominant uncertainties affects the Standard Model predictions are discussed. Recent experimental results include improved determinations of $|V_{cb}|$ by OPAL and CLEO, the confirmation of rare (charmless hadronic and electromagnetic penguin) $B$ decays and preliminary $\sin 2 \beta$ measurements by BaBar and Belle (new world average: $\sin 2 \beta = 0.48 _{-0.24}^{+0.22}$). The updated constraints lead to the Standard Model predictions $\sin 2 \beta = 0.68 \pm 0.03|_{exp}\pm 0.04|_{th} $, $\sin 2 \alpha = -0.43 \pm 0.15 |_{exp} \pm 0.20|_{th} $, $ \gamma = (56 \pm 5|_{exp}\ _{-6}^{+5}|_{th})^{\circ}$, $ \Delta m_{B_{s}} = 16.2 _{-0.3}^{+2.7}|_{exp}\ _{-1.0}^{+1.5}|_{th}\ ps^{-1}$.
\vfill
\end{titlepage}

\section{Present knowledge of the CKM matrix}

A review of the current Standard Model (SM) determination of the CKM matrix and of the unitarity triangle (UT) is presented. Table \ref{Vckm} lists the relevant experimental sources, the best values of the CKM matrix elements of the first two rows and the indirect constraints on the third row, the latter presupposing the SM description of the effective flavour changing neutral-current (FCNC) processes. Experimental values and theoretical estimates used in the analysis of the UT are summarized in Table \ref{constraints} (see Ref. \cite{R1} for a more detailed description).

\begin{table}[htb]
\begin{center}
\caption{Experimental determination of the CKM matrix.}
\label{Vckm}
\renewcommand{\arraystretch}{1.2}
\footnotesize
\begin{tabular}{@{}clcc@{}}
\hline\hline
$|V_{ij}|$ \scriptsize{etc.}   & \scriptsize{from}  &  \scriptsize{value}  &  \scriptsize{ref.} \\
\hline
$G_{F}$   & \it{muon lifetime}  & $1.16639(1)\cdot 10^{-5}GeV^{-2}(\hbar c)^{3}$ &   \cite{PDG} \\
$|V_{ud}|$  & \it{nuclear super-allowed decays}  & $0.9740\pm0.0001_{exp}\pm0.0010_{th}$   &  \cite{nucl,PDG}  \\
$|V_{ud}|$  & \it{neutron decay} & $0.9738\pm0.0016_{exp}\pm0.0004_{th}$ &(\emph{a}) \\
$|V_{ud}|$  & \it{pion $\beta$ decay} & $0.9670\pm0.0160_{exp}\pm0.0008_{th}$ & \cite{nucl} \\
\hline
$|V_{us}|$  & \it{$K_{e3}$ decays} & $0.2200\pm0.0017_{exp}\pm0.0018_{th}$ & \cite{R1} \\
$|V_{us}|$  & \it{hyperon semileptonic decays}  &  0.21 -- 0.24 & (\emph{b}) \\
\hline
   $\begin{array}{c}
            |V_{cd}| \\ |V_{cs}|
         \end{array}$
&  \it{neutrino charm production}
&  $\begin{array}{c}
            0.225\pm0.012 \\ 1.04\pm0.16
         \end{array}$
&  \cite{R1} \\

$|V_{cs}|$  & \it{$D_{e3}$ decays}  &  $1.02\pm0.05_{exp}\pm0.14_{th}$ & (\emph{c}) \\
$|V_{cs}|$  & \it{hadronic $W$ decays} &  $0.99\pm0.02$        & (\emph{d}) \\
\hline

$|V_{ub}|$     & $B \rightarrow \rho \ell \bar{\nu}$ & $(3.25 \pm 0.30_{exp} \pm 0.55_{th}) 10^{-3}$ & \cite{Vub CLEO}\\

$|V_{ub}/V_{cb}|$ &
      \renewcommand{\arraystretch}{1.0}
      \begin{tabular}{@{}l}
           \it {inclusive}  $B \rightarrow X_{u} \ell \bar{\nu}$ \\
           \it{\scriptsize (CLEO, ARGUS)}
      \end{tabular}
& $0.088 \pm 0.006_{exp} \pm 0.007_{th}$ & \scriptsize{text} \\

$|V_{ub}|^{2}$  & \it{inclusive} $b \rightarrow u \ell \bar{\nu}$ \scriptsize{\ (LEP)} & $(16.8 \pm 5.5_{exp} \pm 1.3_{th}) 10^{-6}$ & (\emph{e}) \\

$|V_{cb}|$  & $B \rightarrow D^{\ast} \ell \bar{\nu}$  & $(42.8 \pm 3.3_{exp} \pm 2.1_{th}) 10^{-3}$ & (\emph{f}) \\
$|V_{cb}|$   & \it{inclusive} $b \rightarrow c \ell \bar{\nu}$  & $(41.2 \pm 0.7_{exp} \pm 1.5_{th}) 10^{-3}$ & (\emph{g}) \\

\hline
\scriptsize{$|V_{tb}|^{2}/ \sum_{i}|V_{ti}|^{2}$} & \it{top quark decays} & $0.93_{-0.23}^{+0.31}$ & \cite{Vtb} \\
\hline\hline

\multicolumn{4}{c}{\it effective FCNC processes} \\

\hline

$\begin{array}{c}
            |V_{td}V_{tb}| \\ |V_{ts}/V_{td}|
         \end{array}$
& \it{$B^{0}_{d}/\bar{B}^{0}_{d}$ and $B^{0}_{s}/\bar{B}^{0}_{s}$ oscillations}
&  $\begin{array}{c}
            (8.1\pm0.7_{exp}\pm0.6_{th}) \cdot 10^{-3} \\ > 4.6
         \end{array}$
&  (\emph{h}) \\

$|V_{ts}V_{tb}/V_{cb}|^{2}$ & \it{inclusive} $b \rightarrow s \gamma $  & $0.94\pm0.11_{exp}\pm0.09_{th}$ & (\emph{i})\\
Im($V_{ij}$) & \it {CP-violation measurements:} & \multicolumn{1}{l}{$|\epsilon_{K}| = (2.271 \pm 0.017) \cdot 10^{-3}$} & \cite{PDG}\\
&& \multicolumn{1}{l}{$\epsilon'_{K}/\epsilon_{K} = (19.0 \pm 4.5) \cdot 10^{-4}$} & (\emph{l}) \\
&& \multicolumn{1}{l}{$\sin 2 \beta = 0.48 _{-0.24}^{+0.22}$} & \scriptsize{text}\\

\hline\hline
\end{tabular}
\end{center}
\footnotesize
\vspace{-0.2 cm}
(\emph{a}) from the average experimental values $\tau_{n}= 885.8 \pm 0.8 \ s$ and $g_{A}/g_{V}=-1.2698 \pm 0.0026$ including two recent, precise measurements \cite{Vud neutr} besides those quoted in Ref. \cite{R1}. -- (\emph{b}) While the experimental error is at the $1\%$ level, the spread of values reflects the model dependence of the SU(3)-breaking calculations (Ref. \cite{Vus}). -- (\emph{c}) from the experimental average $\Gamma(D_{e3})=(8.0 \pm 0.8)\cdot 10^{10}\ s^{-1}$ \cite{PDG}, where an exact isospin symmetry between $D^{0} \rightarrow K^{-}e^{+} \nu_{e}$ and $D^{+} \rightarrow \bar{K}^{0}e^{+} \nu_{e}$ is assumed despite the 3\%-level consistency between the two rates (error scaled by 2.2); the theoretical error accounts for the uncertainty in the form-factor normalization. -- (\emph{d}) average of the LEP measurements at centre-of-mass energies up to $202\ GeV$ \cite{Vcs}. -- (\emph{e}) from the LEP average $\mathcal{B}(b \rightarrow u \ell \bar{\nu})=(1.74 \pm 0.57)\cdot 10^{-3}$ \cite{Vub LEP}, using the relation to $|V_{ub}|$ derived in the context of Heavy Quark Theory \cite{HQT}. -- (\emph{f}) from the average $\mathcal{F}(1) \cdot |V_{cb}| = (38.1 \pm 3.0) \cdot 10^{-3}$ (Table \ref{Vcb}), using $\mathcal{F}(1) = 0.89 \pm 0.04$ \cite{HQT}. -- (\emph{g}) The rates measured at the $Z^{0}$ and at the $\Upsilon(4S)$ are \cite{PDG} $\Gamma_{Z^{0}}^{b \rightarrow c \ell \bar{\nu}}=(6.75 \pm 0.14) \cdot 10^{-2}\ ps^{-1}$ and $\Gamma_{\Upsilon(4S)}^{B \rightarrow X_{c} \ell \bar{\nu}}=(6.42 \pm 0.16) \cdot 10^{-2}\ ps^{-1}$ after correction for the $b \rightarrow u$ contribution.
Their weighted average, with an enlarged error, is assumed and the HQT relation to $|V_{cb}|$ \cite{HQT} is used. -- (\emph{h}) $\Delta m_{B_{d}} = 0.487 \pm 0.014 \ ps^{-1} $, $\Delta m_{B_{s}} > 14.9 \ ps^{-1} \ 95\% \ c.l.$ \cite{B osc}; bag factor and decay constants from Table \ref{constraints}. -- (\emph{i}) determined by comparing the experimental average $\mathcal{B}(b \rightarrow s \gamma) =(3.2 \pm 0.4)\cdot 10^{-4}$ (CLEO, ALEPH and Belle \cite{penguin}) with the NLO SM calculation by Chetyrkin \it et al. \rm \cite{pength}. -- (\emph{l}) average of the E731, NA31, KTeV and NA48 results (the latter recently updated) \cite{epsprime}.

\end{table}

\begin{table}[htb]
\begin{center}
\caption{Constraints and input parameters used in the determination of the UT.}
\label{constraints}
\footnotesize 
\renewcommand{\arraystretch}{1.2}
\begin{tabular}{@{}rc@{ \ = \ \ }cll@{}}
\hline\hline

1) & $\lambda$  &  0.2224 & \ $\pm\ 0.0020$ & \scriptsize{$\longleftarrow$ $|V_{us}|$, $|V_{ud}|$, $|V_{cd}|$, $|V_{cs}|$} \\
2) & $|V_{ub}|_{excl}$ \tiny{$(\times 10^{-3})$} &  3.25 & \ $\pm\ 0.30$ & \ $\pm\ \overline{0.55}$   \\
3) & $|V_{ub}/ V_{cb}|$ &  0.088 & \ $\pm\ 0.006$ & \ $\pm\ \overline{0.007}$   \\
4) & $|V_{ub}|^{2}_{incl}$ \tiny{$(\times 10^{-6})$} &  16.8 & \ $\pm\ 5.5$ & \ $\pm\ \overline{1.3}$   \\
5) & $|V_{cb}|_{excl}$ \tiny{$(\times 10^{-3})$} &  42.8 & \ $\pm\ 3.3$ & \ $\pm\ \overline{2.1}$   \\
6) & $|V_{cb}|_{incl}$ \tiny{$(\times 10^{-3})$} &  41.2 & \ $\pm\ 0.7$ & \ $\pm\ \overline{1.5}$ \\
\hline
7) & $|\epsilon_{K}|$ \tiny{$(\times 10^{-3})$} & 2.271 & \ $\pm\ 0.017 $ & \scriptsize $B_{K}=0.94 \pm \overline{0.08}$ \\
8) & $\Delta m_{B_{d}}$ \tiny{$(ps^{-1})$} & 0.487 & \ $\pm\ 0.014 $  &
      \scriptsize $ \left\{ \begin{array}{@{}l}
      f_{B_{d}} = 190 \pm 19 \pm \overline{10}\ MeV \\
      B_{B}=1.30 \pm \overline{0.15}
      \end{array} \right. $ \\
9) & \multicolumn{1}{c@{ \ $>$ \ \ }}{$\Delta m_{B_{s}}$ \tiny{$(ps^{-1})$}} & 14.9 & \ \scriptsize{$95\% \ c.l.$} & \scriptsize 
    $ f_{B_{s}}/f_{B_{d}} = 1.15 \pm \overline{0.04} $ \\
\hline\hline
\end{tabular}
\end{center}
\footnotesize
\vspace{-0.2 cm}
A superscript bar indicates theoretical uncertainties for which flat distributions ($width=2 \cdot error/0.68$) are assumed; the remaining uncertainties are treated as Gaussian errors. The values of $B_{K}$, $B_{B}$ and $f_{B_{s}}/f_{B_{d}}$ are taken from a compilation of lattice-QCD results \cite{R1}, while $f_{B_{d}}$ is an indirect determination obtained from the measured value $f_{D_{s}}=250 \pm 25 \ MeV$ using $ f_{B_{d}}/f_{D_{s}}=0.76 \pm 0.04 $; almost identical estimates are given in recent lattice reviews including the first partially unquenched results \cite{Sach}. The leptonic rate $\mathcal{B}(D_{s}^{+} \rightarrow \mu^{+} \nu_{\mu}) =(0.51 \pm 0.10)\cdot 10^{-2}$, from which the decay constant $f_{D_{s}}$ is extracted, is the average of eight measurements \cite{fDs} of the $\mu^{+} \nu_{\mu}$ and $\tau^{+} \nu_{\tau}$ channels, combined assuming lepton universality and taking into account the correlation due to the uncertainty on the background process $\mathcal{B}(D_{s}^{+} \rightarrow \phi \pi^{+})$.
The parameters $m_{t}=166 \pm 5 \ GeV$, $m_{c}=1.25 \pm \overline{0.7} \ GeV$, $\eta_{cc}=1.38 \pm \overline{0.53}$, $\eta_{ct}=0.47 \pm \overline{0.04}$, $\eta_{tt}=0.574$ (fixed), $\eta_{b}=0.55 \pm \overline{0.01}$ are also used.

\end{table}

\begin{figure}[htb]
\begin{center}
\includegraphics[width=0.60\textwidth]{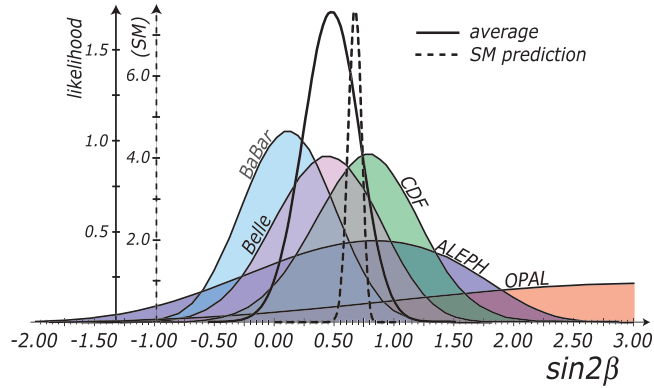}
\caption{P.d.f.'s for $\sin 2 \beta$: measurements, average and SM prediction.}
\label{sin2bfig}
\end{center}
\end{figure}

The first physics results from the asymmetrical $e^{+}e^{-}$ $B$-factories have recently appeared. Preliminary BaBar and Belle measurements of the time dependent CP-asymmetry in $B^{0}/ \bar{B}^{0} \rightarrow J/\psi K_{S}$ decays, combined with a recent update by ALEPH and earlier results by OPAL and CDF \cite{sin2b} (see Table \ref{sin2btab}), lead to the world-average value $\sin 2 \beta = 0.48 _{-0.24}^{+0.22}$, with $0.01 < \sin 2 \beta < 0.91 $ at the 95\% c.l. and a probability of $P(\sin 2 \beta > 0)=98\%$ that CP violation has been measured in this channel. The final probability distribution function (p.d.f.) for $\sin 2 \beta$, computed by combining the likelihood functions obtained in the experiments, is plotted in Figure \ref{sin2bfig}, where the good agreement between the measurements and the SM prediction is made evident. A new measurement of the inclusive $b \rightarrow s \gamma $ branching ratio by Belle confirms previous evaluations by CLEO and ALEPH \cite{penguin}, implying (with $|V_{tb}| \simeq 1$) $|V_{ts}|=(40.2 \pm 3.3) \cdot 10^{-3} $. The hadronic decays $ B \rightarrow K \pi$ and $ B \rightarrow \pi \pi$, which can in principle be related to the CKM angle $\gamma$ even with CP-averaged measurements, have been observed by CLEO, Belle and BaBar \cite{BKpi}.

\begin{table}[htb]
\begin{center}
\caption{$\sin 2 \beta$ measurements and average.}
\label{sin2btab}
\footnotesize
\renewcommand{\arraystretch}{1.3}
\begin{tabular}{lccc}

\hline\hline

\scriptsize{experiment \cite{sin2b}} & $\sin 2 \beta$  & \scriptsize{error from fit}& \scriptsize{other syst.} \\
\hline

OPAL    & 3.2    & $^{+1.8}_{-2.0}$    & $\pm 0.5$  \\
CDF     & 0.79   & $^{+0.41}_{-0.44}$  &     -      \\
ALEPH   & 0.84   & $^{+0.82}_{-1.04}$  & $\pm 0.16$ \\
BaBar   & 0.12   & $\pm 0.37$          & $\pm 0.09$ \\
Belle   & 0.45   & $^{+0.43}_{-0.44}$  & $^{+0.07}_{-0.09}$ \\
\hline
average  & \multicolumn{3}{c}{$0.48 ^{+0.22}_{-0.24}$} \\

\hline\hline
\end{tabular}
\end{center}
\footnotesize
\vspace{-0.2 cm}
To combine the results, the convolution of the original likelihood function with a Gaussian distribution describing the additional systematic error (last column) has been assumed for each experiment.
\end{table}

Other recent B physics results include updated OPAL and CLEO measurements of $|V_{cb}|$ using $B \rightarrow D^{\ast} \ell \bar{\nu}$ decays, which confirm a systematical increase in the central value with respect to previous analyses (see Table \ref{Vcb}), at least partially due to the adoption of an improved parametrization of the form factor $\mathcal{F}(w)$ near the point of zero recoil. Precise measurements of the neutron decay lifetime and axial coupling constant, higher-energy results for $W \rightarrow (hadrons)$ decays at LEP, new analyses of pure leptonic $D_{s}^{\pm}$ decays (WA92, ALEPH) and updated averages by the LEP Working Groups (see Tables \ref{Vckm}, \ref{constraints} and captions) complete the definition of a complex framework of experimental information, which finds a coherent and satisfactory interpretation in the SM description of the processes governed by the CKM matrix.

\begin{table}[htb]
\caption{Determination of $\mathcal{F}(1)|V_{cb}|$ from $B \rightarrow D^{\ast} \ell \bar{\nu}.$}
\label{Vcb}
\centering
\footnotesize 
\renewcommand{\arraystretch}{1.1}
\begin{tabular}{lcc}

\hline\hline
\scriptsize{experiment} \cite{Vcb} & $\mathcal{F}(1) \cdot |V_{cb}|$ \scriptsize{$(\times 10^{-3})$} & \scriptsize{(previous result)} \\
\hline

ALEPH           & $31.9 \pm 1.8  \pm 1.9$     &         -            \\
DELPHI          & $37.95\pm 1.34 \pm 1.59$    & $(35.4\pm 1.9 \pm 2.4)$\\
OPAL            & $37.1 \pm 1.0  \pm 2.0$     & $(32.8\pm 1.9 \pm 2.2)$\\
\hline
\scriptsize{\it LEP average} & $36.1 \pm 1.8$ &                   \\
\hline
CLEO            & $42.4 \pm 1.8  \pm 1.9$     & $(35.1\pm 1.9 \pm 1.9)$\\
\hline
\it{average}   & $38.1 \pm 3.0 $             &                     \\
\hline\hline
\end{tabular}
\end{table}

\begin{figure}[htb]
\begin{center}
\includegraphics[width=0.70\textwidth]{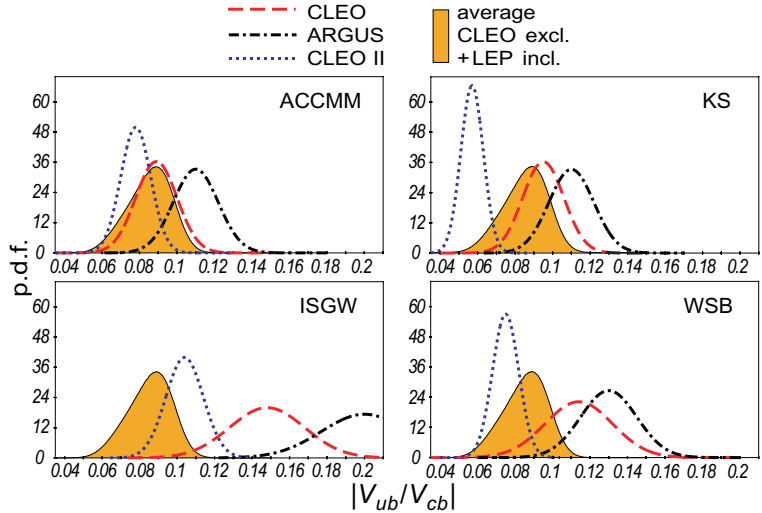}
\caption{Comparison between the lepton end-point measurement of $|V_{ub}/V_{cb}|$ in four models and the combined LEP(inclusive)-CLEO(exclusive) result.}
\label{Vubcbfig}
\end{center}
\end{figure}

Inclusive and exclusive measurements of $B$ semileptonic decays concur to the determination of the crucial constraints $|V_{ub}|$ and $|V_{cb}|$. The results obtained using different techniques (Table \ref{Vckm}) agree with each other within the experimental errors, even if still debated theoretical assumptions are involved (the quark-hadron duality in the inclusive determinations) and cases of strong model dependence exist (the $|V_{ub}|$ exclusive measurement). The most precise measurements involving $b \rightarrow u$ transitions were obtained in the lepton end-point inclusive analyses by CLEO, ARGUS and CLEO II. Model dependent values of $|V_{ub}/V_{cb}|$ were extracted from these results (see Table \ref{VubVcb}); however, the ACCMM model is the only one yielding three consistent values (the KS model was already ruled out in the exclusive analysis by CLEO \cite{Vub CLEO} as predicting two incompatible values of $|V_{ub}|$ from the $\rho \ell \bar{\nu}$ and $\pi \ell \bar{\nu}$ channels). Retaining the average experimental value obtained within the ACCMM model as the best determination and adding a further 8\% `theoretical' error on the basis of the residual discrepancy between the three measurements (error scaled by $\sqrt{\chi^{2}/2} \simeq 1.6$), one gets $|V_{ub}/V_{cb}|=0.088 \pm 0.006_{exp} \pm 0.007_{th}$. This choice is supported a posteriori by the agreement of all three ACCMM predictions with the combination of the independent CLEO (exclusive) and LEP (inclusive) measurements (see Figure \ref{Vubcbfig}).

\begin{table}[htb]
\begin{center}
\caption{Determination of $|V_{ub}/ V_{cb}|$ from inclusive end-point measurements.}
\label{VubVcb}
\footnotesize 
\renewcommand{\arraystretch}{1.1}
\begin{tabular}{@{}cccccc@{}}

\hline\hline
(*)  &  $p_{\ell}$ \scriptsize{(\emph{GeV})} & KS & WSB & ACCMM & ISGW   \\
\hline

CLEO         &
      \renewcommand{\arraystretch}{1.0}
      \begin{tabular}{c}
           \scriptsize{2.2 -- 2.4} \\ \scriptsize{2.4 -- 2.6}
      \end{tabular}
     & $0.095\pm0.011$ & $0.114\pm0.018$ & $0.089\pm0.011$ 
&  $0.148\pm0.020$ \\
ARGUS        &  2.3 -- 2.6
     & $0.110\pm0.012$ & $0.130\pm0.015$ & $0.110\pm0.012$ 
&  $0.200\pm0.023$ \\
CLEO II      &
      \renewcommand{\arraystretch}{1.0}
      \begin{tabular}{c}
           \scriptsize{2.3 -- 2.4} \\ \scriptsize{2.4 -- 2.6}
      \end{tabular}
     & $0.057\pm0.006$ & $0.075\pm0.007$ & $0.078\pm0.008$ 
&  $0.104\pm0.010$ \\
\hline
\multicolumn{2}{c}{\it{average}} 
     & $0.073\pm0.005$ & $0.088\pm0.006$ & $0.088\pm0.006$ 
&  $0.124\pm0.008$ \\
\multicolumn{2}{c}{$\chi^{2}$ (c.l.)}
     & 20.6 (0.003\%)  & 13.4 (0.1\%)    & 4.9 (8.6\%)     &  16.4 (0.03\%)\\
\hline\hline
\end{tabular} \\
\vspace{0.1 cm}
(*) see Refs. \cite{Vubcb exp} (experimental results) and \cite{Vubcb mod} (models).
\end{center}
\end{table}

The constraints listed in Table \ref{constraints} have been used as the input values of a Bayesian determination of the UT parameters, realized through a Monte Carlo scanning of all the uncertain quantities within the allowed ranges, assuming flat and Gaussian prior distributions. The Wolfenstein parameter $\lambda$ has been determined from a fit to the available measurements of $|V_{us}|$, $|V_{cd}|$ ($\simeq \lambda$), $|V_{ud}|$ ($\simeq 1-\lambda^{2}/2-\lambda^{4}/8$) and $|V_{cs}|$ ($\simeq 1-\lambda^{2}/2-\lambda^{4}(1/8+A^{2}/2)$) (the result, $\lambda=0.2224 \pm 0.0020$, is one standard deviation higher than the direct measurement of $|V_{us}|$.
The 68\% c.l. results for the co-ordinates $(\bar{\rho}, \bar{\eta})$ of the vertex of the UT, the CP-violation observables $\sin 2 \beta$, $\gamma$ and $\sin 2 \alpha$ and the $B_{s}$ oscillation parameter $\Delta m_{B_{s}}$ are
\begin{center}
$\begin{array}{r@{\ =\ }c@{\ =\ }c}
\bar{\rho} & 0.21 \pm 0.03|_{exp}\pm 0.04|_{th} & 0.21 \pm 0.05 \\
\bar{\eta} & 0.31 \pm 0.02|_{exp}\pm 0.03|_{th} & 0.31 \pm 0.03 \\
\sin 2 \beta  &  0.68 \pm 0.03|_{exp}\pm 0.04|_{th}   &  0.68 \pm 0.05 \\
\sin 2 \alpha & -0.43 \pm 0.15 |_{exp} \pm 0.20|_{th} & -0.43 \pm 0.25 \\
\gamma &(56 \pm 5|_{exp}\ _{-6}^{+5}|_{th})^{\circ} & (56 _{-8}^{+7})^{\circ}\\
\Delta m_{B_{s}} & 16.2 _{-0.3}^{+2.7}|_{exp}\ _{-1.0}^{+1.5}|_{th}\ ps^{-1} & 16.2_{-1.0}^{+3.1}\ ps^{-1} \\
\end{array}$
\end{center}

\begin{figure}[htb]
\begin{center}
\includegraphics[width=0.90\textwidth]{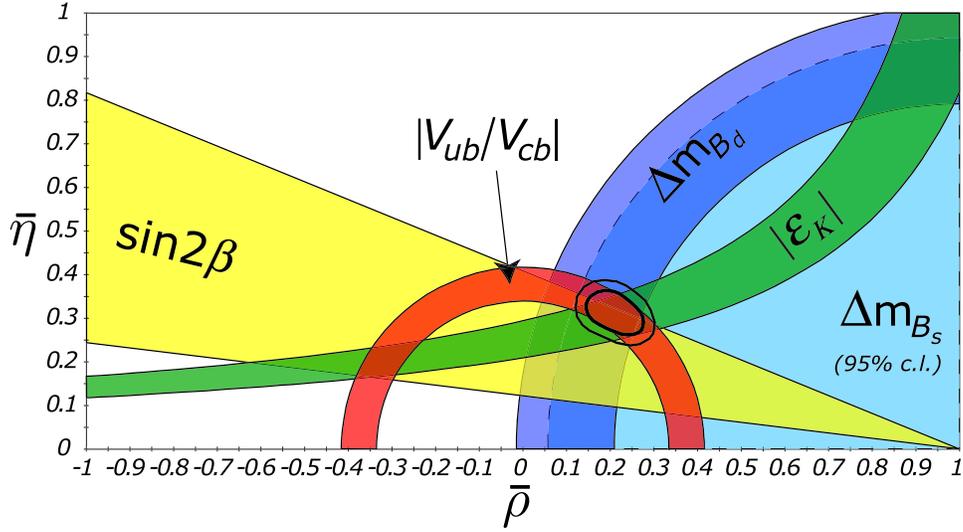}
\caption{Constraints on the UT ($\pm 1\sigma$) and allowed region (68 and 95\% c.l.) for the vertex $(\bar{\rho}, \bar{\eta})$. The lines corresponding to the present experimental average of $\sin2 \beta$ are also shown.}
\label{rhoeta}
\end{center}
\end{figure}

\begin{figure}[htb]
\begin{center}
\includegraphics[width=0.60\textwidth]{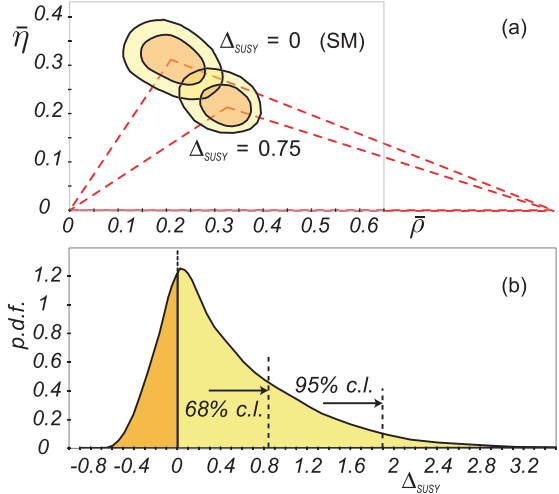}
\caption{SM and MSSM profiles of the unitarity triangle and p.d.f. for $\Delta_{SUSY}$.}
\label{SUSY}
\end{center}
\end{figure}

The region selected by the constraints in the $\bar{\rho}, \bar{\eta}$ plane is shown in Figure \ref{rhoeta}. Even if some of the parameters are known to a precision of better than 10\%, the current profile of the UT cannot be considered as a precise determination: this is especially true if one believes that new physics will manifest itself through \emph{small} deviations from the SM, as is suggested by the consistency of the present constraints. As an example, Figure \ref{SUSY} (a) shows how the SM profile of the UT can be modified according to a class of minimal supersymmetric extensions (MSSM), in which no new phases enter into the expressions of the CP-violating asymmetries of B decays and the analysis of the UT can be performed in a similar way as in the SM (see Ref. \cite{susyb} for a detailed review and further examples). The contribution of new physics in this class of models is measured by a single additional parameter $ \Delta_{SUSY} \geq 0 $, entering into the expressions of $\Delta m_{B_{d}}$, $\Delta m_{B_{s}}$ and $|\epsilon_{K}|$ as a sort of `correction' to the top-quark loop functions: the substitution $S(m_{t}^{2}/m_{W}^{2}) \rightarrow (1+ \Delta_{SUSY})\cdot S(m_{t}^{2}/m_{W}^{2})$ has to be made, with $\Delta_{SUSY}=0$ in the SM. There is a significant overlap between the UT profile in the SM and the one corresponding to the maximum deviation from the SM ($\Delta_{SUSY} = 0.75$, maximum value allowed by limits on supersymmetric masses and electric dipole moments \cite{susyb}). In particular, a precise measurement of $\sin 2 \beta $ would not be able to exclude one or the other, the MSSM prediction being included between 0.46 and 0.70 at the 95\% c.l.; there may be an additional difficulty in interpreting the result, if the unknown new physics should also affect the \emph{measurement} of $\sin 2 \beta$ (i.e. its relation with the measured CP-asymmetry). This is one reason why the \emph{sides} of the UT will have to be known with higher precision independently of the direct measurements of the angles. Extensive lattice-QCD simulations and an enhanced theoretical understanding of $B$ decays at the quark-level are first priorities to this purpose; at the same time, new measurements will provide additional constraints (e.g. $|V_{ub}/V_{td}|$ from $\mathcal{B}(B^{+}\rightarrow \mu^{+} \nu_{\mu})/\Delta m_{B_{d}}$ \cite{Ros}) and an experimental check of the theoretical estimates ($f_{B_{d}}$, $f_{D_{s}}$, $f_{D}$).

\section{Over-consistency and theoretical errors}

There is an interesting coincidence related to the example of the MSSM models. When the parameter $\Delta_{SUSY}$ is left free to vary and determined from the UT constraints, the p.d.f. plotted in Figure \ref{SUSY} (b), showing a sharp maximum just in $\Delta_{SUSY} = 0$, is obtained. This result, though not precise enough to further restrict the allowed range of values, is an example of how difficult is to detect small incongruities in the present SM determination of the UT, even when constraints crucially related to the possible contribution of new physics are concerned.

A rough quantification of the consistency between the constraints is given by the result of a $\chi^{2}$ test, carried out after adding in quadrature experimental and theoretical errors: the minimum value is $\chi^{2} = 1.8$ with 5 degrees of freedom, corresponding to an 88\% probability of finding less consistent results.

It may be argued that the results of a statistical analysis of the UT should be regarded with a certain reserve due to the large theoretical errors entering into the constraints. Actually, a declaredly `conservative' treatment of the theoretical uncertainties is adopted in other analyses of the UT \cite{Art,Ros}, where the need for a confirmation of some crucial assumptions (the quark-hadron duality, the reliability of quenched or partially unquenched lattice-QCD calculations) and the difficulty of establishing the meaning and the magnitude of the non-statistical errors are emphasized.

On the other hand, one aspect of the extraordinary consistency of the current SM constraints on the UT is that the dependence of the results of a `global fit' on the choice of the theoretical errors is rather weak, as is shown by the following examples (the results are listed in Table \ref{errors}, where the label `standard' refers to the results obtained with the original choice of flat theoretical errors, as indicated in Table \ref{constraints}):

i) almost identical results are obtained when all the errors, including the `theoretical' ones, are assumed to be Gaussian;

ii) both when all the theoretical errors are set equal to zero and when they are multiplied respectively by factors of 2 and 4, all the mean values remain practically unchanged.

\begin{table}[htb]
\begin{center}
\caption{Dependence of the results on the choice of the theoretical errors.}
\label{errors}
\footnotesize
\renewcommand{\arraystretch}{1.1}
\begin{tabular}{@{}lcccc@{}}

\hline\hline

 &  $\bar{\rho}$  &  $\bar{\eta}$  &  $\sin 2 \beta$  &  $\gamma\ (^{\circ})$ \\
\hline

\scriptsize{th. errors $=0$}
         & $0.21 \pm 0.03$ & $0.30 \pm 0.02$ & $0.67 \pm 0.03$ & $56 \pm 5$  \\
\scriptsize{Gaussian errors}
         & $0.21 \pm 0.05$ & $0.31 \pm 0.04$ & $0.68 \pm 0.05$ & $56 \pm 8$  \\
\hline
\scriptsize{(\emph{standard})}
         & $0.21 \pm 0.05$ & $0.31 \pm 0.03$ & $0.68 \pm 0.05$ & $56 \pm 8$\\
\hline
\scriptsize{th. errors $\times 2$}
         & $0.22 \pm 0.08$ & $0.32 \pm 0.06$ & $0.70 \pm 0.09$ & $56 \pm 12$\\
\scriptsize{th. errors $\times 4$}
         & $0.22 \pm 0.13$ & $0.33 \pm 0.10$ & $0.72 \pm 0.15$ & $57 \pm 20$\\
\hline\hline
\end{tabular} \\
\vspace{0.1 cm}
Mean and standard deviation of the final p.d.f. are given for each result.
\end{center}
\end{table}
 
Similar and complementary examples are presented in Ref. \cite{AS}, confirming in particular that the \emph{shape} of the prior distribution assumed for the parameters affected by theoretical uncertainty is not relevant. Moreover, the mean values of the predictions are independent of \emph{how large} the theoretical errors are chosen. As a further example, the (above-mentioned) low chi-square value may indicate that the current estimates of the theoretical uncertainties, in accordance with their non-statistical nature, tend to be associated with confidence levels higher than the assumed one-standard-deviation value.

Apparently, the present determination of the UT is a configuration of extraordinary stability; enhanced theoretical calculations, while essential to improve the existing constraints, are unlikely to upset this picture.

\subsection*{Conclusions}

Improved measurements and new results confirm a high degree of coherence in the SM description of the processes governed by the CKM matrix. In particular, the present determination of the UT is extraordinarily self-consistent and only weakly dependent on definite assumptions about the meaning and the magnitude of the theoretical errors; nevertheless, it is not precise enough to put stringent constraints on at least certain minimal extensions of the SM. The experiments are facing a real challenge in their attempt to cause possible signals of new physics to emerge from this picture. For this purpose, a better determination of the sides of the UT, involving the progress of both the experimental and the theoretical techniques, should be given the same high priority as the direct measurements of the angles.

\subsection*{Acknowledgements}

I am indebted to Peter Schlein and to the organizers of this conference for giving me the opportunity of this presentation. This report is part of the activity of the Bologna University and INFN Group engaged in the HERA-B experiment, whose suggestions and constant support are gratefully acknowledged.


\begin{thebibliography}{99}

\itemsep = 0 cm

\bibitem{R1} M. Bargiotti et al., La Rivista del Nuovo Cimento, Vol. 23,
             N.3 (2000) 1, hep-ph/0001293.

\bibitem{PDG}   D.E. Groom et al. (Particle Data Group),
                Eur. Phys. J. C15 (2000) 1.

\bibitem{sin2b} K. Ackerstaff et al. (OPAL), Eur. Phys. J. C 5 (1998) 379;
                T. Affolder et al. (CDF), Phys. Rev. D 61 (2000) 072005;
                B. Aubert et al. (BaBar), hep-ex/0008048;
                H. Aihara for the Belle Collaboration, hep-ex/0010008;
                R. Barate et al. (ALEPH), hep-ex/0009058.

\bibitem{BKpi} D. Cronin-Hennessy et al. (CLEO), hep-ex/0001010;
               A. Abashian et al., BELLE-CONF-0005, ICHEP 2000;
               C. Aubert et al. (BaBar), hep-ex/0008057.

\bibitem{nucl}  J.C. Hardy and I.S. Towner, nucl-th/9809087.

\bibitem{Vud neutr} S. Arzumanov et al., Nucl. Instr. Meth. A 440 (2000) 511;
                    J. Reich et al., Nucl. Instr. Meth. A 440 (2000) 535.

\bibitem{Vus}   R. Flores-Mendieta et al., Phys. Rev. D 54 (1996) 6855.

\bibitem{Vcs} ALEPH Collaboration, ALEPH 2000-005, ICHEP 2000;
              P. Buschmann et al., DELPHI 2000-140 CONF 439, ICHEP 2000;
              L3 Collaboration, L3 Note 2514;
              G. Abbiendi et al. (OPAL), Phys. Lett. B490 (2000) 71.

\bibitem{Vcb} D. Buskulic et al. (ALEPH), Phys. Lett. B 395 (1997) 373;
              P. Abreu et al. (DELPHI), Z. Phys. C 71 (1996) 539;
              DELPHI Collaboration, DELPHI 99-107 CONF 294, EPS-HEP 1999;
              K. Ackerstaff et al. (OPAL), Phys. Lett. B 395 (1997) 128;
              G. Abbiendi et al. (OPAL), hep-ex/0003013;
              B. Barish et al. (CLEO), Phys. Rev. D 51 (1995) 1014;
              J.P. Alexander et al. (CLEO), hep-ex/0007052.

\bibitem{HQT} I.I. Bigi, hep-ph/9907270; N. Uraltsev, hep-ph/9905520;
              A.H. Hoang, Z. Ligeti and A.V. Manohar,
              Phys. Rev. Lett. 82 (1999) 277. 

\bibitem{Vub CLEO} J.P. Alexander et al. (CLEO), Phys. Rev. Lett. 77 (1996)
                   5000; B.H. Behrens et al. (CLEO), hep-ex/9905056.

\bibitem{Vub LEP} LEP Vub Working Group, LEPVUB-00/01.

\bibitem{Vubcb exp} R. Fulton et al. (CLEO), Phys. Rev. Lett. 64 (1990) 16;
                    H. Albrecht et al. (ARGUS), Phys. Lett. B 255 (1991) 297;
                   J. Bartelt et al. (CLEO II), Phys. Rev. Lett. 71 (1993) 4111.

\bibitem{Vubcb mod} J.G. K\"orner and G.A. Schuler, Z. Phys. C 38 (1988) 511;
                    M. Wirbel, B. Stech and M. Bauer, Z. Phys. C 29 (1985) 637;
                    G. Altarelli, N. Cabibbo, G. Corb\`o, L. Maiani and
                    G. Martinelli, Nucl. Phys. B 208 (1982) 365;
                    N. Isgur, D. Scora, B. Grinstein and M.B. Wise,
                    Phys. Rev. D 39 (1989) 799.

\bibitem{Vtb} CDF Collaboration, CDF/PUB/TOP/CDFR/5028.

\bibitem{B osc} LEP B-oscillation Working Group, Results for the Summer 2000
                conferences \\ (http://lepbosc.web.cern.ch/LEPBOSC/).

\bibitem{penguin} S. Glenn et al., CLEO CONF 98-17, ICHEP 1998;
                  R. Barate et al. (ALEPH), Phys. Lett. B 429 (1998) 169;
                  A. Abashian et al., BELLE-CONF-0003, ICHEP 2000.

\bibitem{pength} K. Chetyrkin et al., Phys. Lett. B 400 (1997) 206.

\bibitem{epsprime} L.K. Gibbons et al. (E731), Phys. Rev. Lett. 70 (1993) 1203;
                   G.D. Barr et al. (NA31), Phys. Lett. B 317 (1993) 233;
                   A. Alavi-Harati et al. (KTeV),
                   Phys. Rev. Lett. 83 (1999) 22;
                   C. Biino (NA48), these Proceedings.

\bibitem{fDs} S. Aoki et al. (WA75), Prog. Theor. Phys. 89 (1993) 131;
              J.Z. Bai et al. (BES), Phys. Rev. Lett. 74 (1995) 4599;
              K. Kodama et al. (E653), Phys. Lett. B 382 (1996) 299;
              M. Acciarri et al. (L3), Phys. Lett. B 396 (1997) 327;
              F. Parodi et al. (DELPHI), DELPHI 97-105 CONF 87, EPS-HEP 1997;
              M. Chada et al. (CLEO), Phys. Rev. D 58 (1998) 032002;
              Y. Alexandrov et al. (BEATRICE), Phys. Lett. B 478 (2000) 31;
              ALEPH Collaboration, ALEPH 2000-062 CONF 2000-041, ICHEP 2000.

\bibitem{Sach} C.T. Sachrajda, these Proceedings.

\bibitem{susyb} A. Ali and D. London, hep-ph/0002167.

\bibitem{Ros} J. Rosner, these Proceedings.

\bibitem{Art} M. Artuso, these Proceedings.

\bibitem{AS} A. Stocchi, these Proceedings.

\end{thebibliography}
\end{document}